\documentclass{emulateapj}

\shorttitle{Outflows from dSphs}
\shortauthors{Kirby, Martin, \& Finlator}
\slugcomment{Accepted to ApJL, 2011 Oct 25}

\begin{document}

\title{Metals Removed by Outflows from Milky Way Dwarf Spheroidal
  Galaxies\altaffilmark{*}}

\author{Evan~N.~Kirby\altaffilmark{1,2},
  Crystal~L.~Martin\altaffilmark{3}, and Kristian
  Finlator\altaffilmark{2,3}}

\altaffiltext{*}{The data presented herein were obtained at the
  W.M. Keck Observatory, which is operated as a scientific partnership
  among the California Institute of Technology, the University of
  California and the National Aeronautics and Space
  Administration. The Observatory was made possible by the generous
  financial support of the W.M. Keck Foundation.}
\altaffiltext{1}{California Institute of Technology, 1200
  E.\ California Blvd., MC 249-17, Pasadena, CA 91125, USA}
\altaffiltext{2}{Hubble Fellow.}  \altaffiltext{3}{Department of
  Physics, University of California, Santa Barbara, CA 93106, USA}



\begin{abstract}

The stars in the dwarf spheroidal satellite galaxies (dSphs) of the
Milky Way are significantly more metal-poor than would be expected
from a closed box model of chemical evolution.  Gas outflows likely
carried away most of the metals produced by the dSphs.  Based on
previous Keck/DEIMOS observations and models, we calculate the mass in
Mg, Si, Ca, and Fe expelled from each of eight dSphs.  Essentially,
these masses are the differences between the observed amount of metals
present in the dSphs' stars today and the inferred amount of metals
produced by supernovae.  We conclude that the dSphs lost 96\% to
$>99\%$ of the metals their stars manufactured.  We apply the observed
mass function of Milky Way dSphs to the ejected mass function to
determine that a single large dSph, like Fornax, lost more metals over
10~Gyr than all smaller dSphs combined.  Therefore, small galaxies
like dSphs are not significant contributors to the metal content of
the intergalactic medium.  Finally, we compare our ejected mass
function to previous X-ray measurements of the metal content of the
winds from the post-starburst dwarf irregular galaxy NGC~1569.
Remarkably, the most recent starburst in that galaxy falls exactly on
the ejected mass-stellar mass relation defined by the Milky Way dSphs.

\end{abstract}


\section{Introduction}
\label{sec:intro}

In a closed box model of chemical evolution \citep{van62,sch63}, the
peak of the stellar metallicity distribution lies at the
nucleosynthetic yield.  For most supernovae (SNe), the nucleosynthetic
yield is about solar metallicity or higher \citep{woo95,lim03,nom06}.
However, the peaks of the stellar metallicity distributions of all
known dwarf spheroidal galaxies (dSphs) are less than solar
\citep[e.g.,][]{aar85,buo85}. The dSphs likely lost metals during star
formation, which reduced their effective nucleosynthetic yields
\citep{har76,yos87}.  Almost no dSph with 250~kpc of the Milky Way
(MW) contains any detectable gas at present
\citep[e.g.,][]{bli00,grc09}.  Therefore, the missing metals are not
hiding in the galaxies' interstellar medium.  The metals have been
lost.

Supernova winds drive gas out of galaxies, particularly small ones
\citep{lar74,dek86}.  It is possible to detect this gas flowing out of
galaxies, but it is very difficult to measure the metallicity or even
the mass of galactic winds.  Absorption line studies of distant
galaxies, including galaxies at cosmological redshift, can measure
column densities and outflow velocities
\citep{mar05,rup05,wei09,ste10,rub10} but not mass or composition.
Smaller galaxies may be even more prone to gas outflows due to their
shallow gravitational potential wells.  \citet{mar99} showed that the
observed hot gas around gas-rich dwarf irregular galaxies (dIrrs) in
the nearby universe must be escaping from those galaxies in the form
of outflows.  Just a few such galaxies are suitable for X-ray
spectroscopy to measure the composition of the escaping gas
\citep[e.g., NGC~1569][]{mar02},

Dwarf galaxies, which are numerous and have shallow gravitational
potential wells, may be important contributors to the metal content of
the the intergalactic medium \citep{opp08,mar10} or to their host
galaxies.  However, most nearby dwarf galaxies, especially dSphs, have
not experienced appreciable star formation for several Gyr.  As a
result, it is even more difficult to observe gas expelled from Local
Group dwarfs because it must have happened so long ago.

\citet{kir10,kir11b} recently measured the stellar metal content of
dSphs.  The amount of metals produced over cosmic time for an
individual dSph may be estimated from its stellar mass, an assumed
stellar initial mass function (IMF), and theoretical SN yields and
explosion rates.  The difference between the inferred metal production
by SNe and the current stellar metal content gives the amount of
metals the galaxy has lost.  This letter presents that calculation and
discusses the role of dSphs in enriching the intergalactic medium
(IGM) and their host galaxies.


\section{Calculations of Metals Expelled}
\label{sec:obs}

\citet[][hereafter K11a]{kir11a} constructed a basic chemical
evolution model in order to explain the metallicity and [$\alpha$/Fe]
distributions of eight MW dSphs: Fornax, Leo~I, Leo~II, Sculptor,
Sextans, Draco, Canes Venatici~I, and Ursa Minor.  The model assumes
that each galaxy begins its life with a certain amount of gas
($M_g(0)$) but no stars.  Primordial, metal-free gas flows into the
galaxy at a prescribed inflow rate ($dM_g/dt = A_{\rm in} t \,
e^{-t/\tau_{\rm in}}$).  Stars form according to a star formation rate
law ($dM_*/dt = A_* M_g^{\alpha}$).  Type~II SNe explode from massive
(10--100~$M_{\sun}$) stars, and Type~Ia SNe explode according to an
empirical delay time distribution \citep{mao10}.  Most importantly for
this letter, the galaxy loses a fixed amount of gas ($A_{\rm out}$)
during each SN explosion.  \citeauthor*{kir11a} varied the parameters
$M_g(0)$, $A_{\rm in}$, $A_*$, $A_{\rm out}$, $\tau_{\rm in}$, and
$\alpha$ until the modeled metallicity and [$\alpha$/Fe] distributions
matched the observed distributions.

This model provides good matches to the abundance distributions of the
less luminous dSphs.  The inferred star formation histories also were
qualitatively consistent with histories measured from color-magnitude
diagrams.  Essentially, nearly all stars in the less luminous dSphs
are ancient, and they formed in 2~Gyr or less.  However, the models
for the most luminous dSphs (Fornax, Leo~I, and to some extent,
Leo~II) have chemically measured star formation durations of only
$\sim 1$~Gyr.  The color-magnitude diagrams for these galaxies rule
out such short star formation lifetimes, and they also rule out single
bursts of star formation.  These dSphs experienced several epochs of
an increase in star formation rate (SFR) followed by a decrease.  The
\citeauthor*{kir11a} models permit only one episode of star formation
(a single rise and fall of SFR).  Therefore, the models are imperfect
descriptions of Fornax, Leo~I, and possibly Leo~II.

On the other hand, for an old enough population, such as the MW dSphs,
most of the SNe that will ever explode have already exploded.
Therefore, the star formation history (SFR as a function of time) has
a small effect on the total metal production integrated over cosmic
time.  Because SN yields are metallicity-dependent, the star formation
history does matter to a small degree.

\begin{deluxetable*}{lcccccc}
\tablewidth{0pt}
\tablecolumns{7}
\tablecaption{Inferred Ejected Masses Over 10~Gyr\label{tab:outflows_10Gyr}}
\tablehead{\colhead{DSph} & \colhead{$L~(L_{\sun})$} & \colhead{$M_*~(M_{\sun})$} & \colhead{$M_{\rm{Mg}}~(M_{\sun})$} & \colhead{$M_{\rm{Si}}~(M_{\sun})$} & \colhead{$M_{\rm{Ca}}~(M_{\sun})$} & \colhead{$M_{\rm{Fe}}~(M_{\sun})$}}
\startdata
Fornax           & $(1.8 \pm 0.5) \times 10^7$ & $(1.9 \pm 0.5) \times 10^7$ & $1.2 \times 10^4$ & $2.9 \times 10^4$ & $1.9 \times 10^3$ & $8.3 \times 10^4$ \\
Leo~I            & $(5.6 \pm 1.7) \times 10^6$ & $(4.6 \pm 1.4) \times 10^6$ & $3.7 \times 10^3$ & $7.9 \times 10^3$ & $4.9 \times 10^2$ & $2.1 \times 10^4$ \\
Sculptor         & $(2.3 \pm 1.1) \times 10^6$ & $(1.2 \pm 0.6) \times 10^6$ & $9.9 \times 10^2$ & $2.1 \times 10^3$ & $1.3 \times 10^2$ & $5.5 \times 10^3$ \\
Leo~II           & $(6.6 \pm 1.9) \times 10^5$ & $(1.4 \pm 0.4) \times 10^6$ & $1.2 \times 10^3$ & $2.7 \times 10^3$ & $1.7 \times 10^2$ & $7.1 \times 10^3$ \\
Sextans          & $(4.1 \pm 1.2) \times 10^5$ & $(8.5 \pm 2.4) \times 10^5$ & $6.6 \times 10^2$ & $1.5 \times 10^3$ & $9.2 \times 10^1$ & $3.8 \times 10^3$ \\
Draco            & $(2.7 \pm 0.4) \times 10^5$ & $(9.1 \pm 1.4) \times 10^5$ & $7.0 \times 10^2$ & $1.6 \times 10^3$ & $9.7 \times 10^1$ & $4.0 \times 10^3$ \\
Can.\ Ven.~I     & $(2.3 \pm 0.4) \times 10^5$ & $(6.3 \pm 1.1) \times 10^5$ & $4.9 \times 10^2$ & $1.1 \times 10^3$ & $6.8 \times 10^1$ & $2.8 \times 10^3$ \\
Ursa Minor       & $(2.2 \pm 0.7) \times 10^5$ & $(5.6 \pm 1.7) \times 10^5$ & $3.9 \times 10^2$ & $9.1 \times 10^2$ & $5.5 \times 10^1$ & $2.3 \times 10^3$ \\
\enddata
\tablerefs{$L$ (luminosity): \citet{mar08} for Canes Venatici~I; \citet{irw95} otherwise.  $M_*$ (stellar mass): \citet{woo08}, except that we assumed that Canes Venatici~I the same $M_*/L$ as Ursa Minor.}
\end{deluxetable*}

As a supplement to the work of \citeauthor*{kir11a}, we compute the
mass in a number of metals, $M_{\rm ejected}$, that the dSphs lost to
SN winds.  The \citeauthor*{kir11a} chemical evolution model tracks
the amount and composition of expelled gas with each 1~Myr time step.
The model was terminated when the dSphs were exhausted of gas.
Although dSphs lose a small fraction of stars to tidal interactions
with the MW \citep{maj00}, we neglect this deduction from the final
stellar mass.  The existence of the luminosity-metallicity relation
\citep{mat98,kir11b} limits the severity of tidal stripping of
luminous matter \citep[also see][]{pen08} because tidally stripped
versions of massive satellites would be just as metal-rich as other
massive satellites, but small, metal-rich satellites do not exist.

In order to mitigate the uncertainties imposed on $M_{\rm ejected}$
from uncertain star formation histories, we advance the chemical
evolution model for each dSph to an age of 10~Gyr.  We compute the
amount of ejecta---mostly in Fe---from Type~Ia SNe over 10~Gyr, even
though star formation ends in the models long before then.  Like
\citeauthor*{kir11a}, we use \citeauthor{nom06}'s (\citeyear{nom06})
nucleosynthetic yields for Type~II SNe, \citeauthor{iwa99}'s
(\citeyear{iwa99}) yields for Type~Ia SNe, and \citeauthor{mao10}'s
(\citeyear{mao10}) delay time distribution for Type~Ia SNe.  The first
Type~Ia SN explodes 100~Myr after the birth of its progenitor, and the
Type~Ia SN rate declines as $t^{-1.1}$.  Based on the theoretical
yields and IMF \citep{kro93} we have adopted, Type~Ia SNe produce 94\%
of the iron over 10~Gyr.  Asymptotic giant branch stars are included
in the chemical evolution model, but at these low metallicities, they
are responsible for only a few percent of the total production of Mg,
Si, Ca, and Fe \citep{kar10}.

Table~\ref{tab:outflows_10Gyr} gives $M_{\rm ejected}$ for Mg, Si, Ca,
and Fe for the eight MW dSphs in our sample.  We do not attempt to
estimate errors on $M_{\rm ejected}$ because the dominant source of
uncertainty is the SN yields.  Differences between different yield
models \citep[e.g.,][]{woo95,nom06} can be a factor of several for
some elements.

If Type~Ia SNe have a prompt component \citep{sca05}, then the Type~Ia
SNe at early times could feed the same wind as the Type~II SNe.
However, the delayed Type~Ia SNe at late times explode at a low rate.
Without continuous explosions, it may not be possible to drive a wind.
Nonetheless, nearly all dSphs are free of gas.  Therefore, the
``missing metals''---the difference between the current metal content
of the dSph and the amount of metals estimated to have been produced
by all of the dSph's SNe---have left the dSphs, whether in coherent
winds or otherwise.

\citeauthor*{kir11a} assumed that the composition of the gas lost to
SN winds was the same as the gas within the galaxy at the time of
loss.  However, outflows could be metal-enhanced, especially because
the SNe that expel gas from the galaxy are the same events that enrich
the galaxy with metals \citep{vad86,mac99}.  The mass of the outflow
is related inversely to its unknown metallicity, and
\citeauthor*{kir11a} showed that increasing the outflowing metallicity
decreased the total amount of galactic mass loss required by as much
as a factor of 40.  Given this uncertainty, we estimate only the mass
in ejected metals and not the total mass of ejected material.  The
uncertainty in the metallicity of the outflows does not extend to the
independent estimate of the ejected metal mass.


\section{Relations with Stellar Mass}
\label{sec:mstar}

\subsection{Total Metal Output}

Figure~\ref{fig:ejecta} shows the mass in each of several metals lost
to SN winds summed over 10~Gyr.  Not surprisingly, the dSphs with
larger stellar masses eject more metals than the dSphs with smaller
stellar masses.  The $M_{\rm ejected}-M_*$ relation is linear, which
is the result expected if the dSphs lost all of their metals.  The
intercepts of the best-fit lines indicate that dSphs lost a constant
0.5\% of their final stellar masses in the form of Fe, 0.2\% as Si,
0.1\% as Mg, and 0.01\% as Ca.

Of course, the dSphs could not have truly lost all of their metals
because the stars are not metal-free.  However, the amount of metals
retained ($M_{\rm retained}$, the filled points in
Figure~\ref{fig:ejecta}) in stars is tiny compared to the amount of
metals produced.  \citet{kir11b} measured the average metallicity of
Fornax to be $\langle\mathrm{[Fe/H]}\rangle = -1.0$.\footnote{We adopt
  the solar abundance of Fe as $12 + \log (n({\rm Fe})/n({\rm H})) =
  7.52$ \citep{sne92}.  We adopt 7.58, 7.55, and 6.36 for Mg, Si, and
  Ca, respectively \citep{and89}.}  Given its stellar mass of $1.9
\times 10^7~M_{\sun}$, Fornax's stars contain $4 \times 10^3~M_{\sun}$
of Fe.  Even though Fornax retained the highest fraction of Fe of all
eight dSphs considered here, the galaxy presently has only 4\% of the
Fe that it produced.  In Ursa Minor, that fraction is only 0.2\%.

The bottom panel of Figure~\ref{fig:ejecta} expresses metal retention
in another way.  The black circles show the stellar mass-iron
metallicity relation for MW dSphs \citep{kir11b}.  The other points
show the mass-metallicity relations for other elements.  The smaller
dSphs retain a larger fraction of $\alpha$ elements relative to iron
than the larger dSphs.  The dependence on stellar mass indicates that
the larger dSphs formed stars for longer and therefore experienced a
higher ratio of Type~Ia to Type~II SNe.

\subsection{Relative Metal Output}

The mass-metallicity relation indicates that the more massive dSphs
retained more of their metals in stars rather than losing them to
galactic winds.  Therefore, the less massive dSphs contribute a larger
fraction of their metal mass to the IGM or to their host galaxies.  To
further explore the importance of small dSphs, we convolve the ejected
mass function with the stellar mass function for MW dSphs.
\citet{kop08} calculated the luminosity function of MW dSphs by
correcting the observed Sloan Digital Sky Survey luminosity function
for sky coverage and bias against low surface brightness dSphs.  The
result was $dN/dM_V = 10 \times 10^{0.1(M_V + 5)}$.  Assuming the
stellar mass-to-light ratio for an old population, $M_*/L = 2$, the
relation becomes $dN/d\log M_* = 284\, M_*^{-0.25}$ where $M_*$ is
measured in $M_{\sun}$.

\begin{figure}[t!]
  \centering
  \includegraphics[width=\columnwidth]{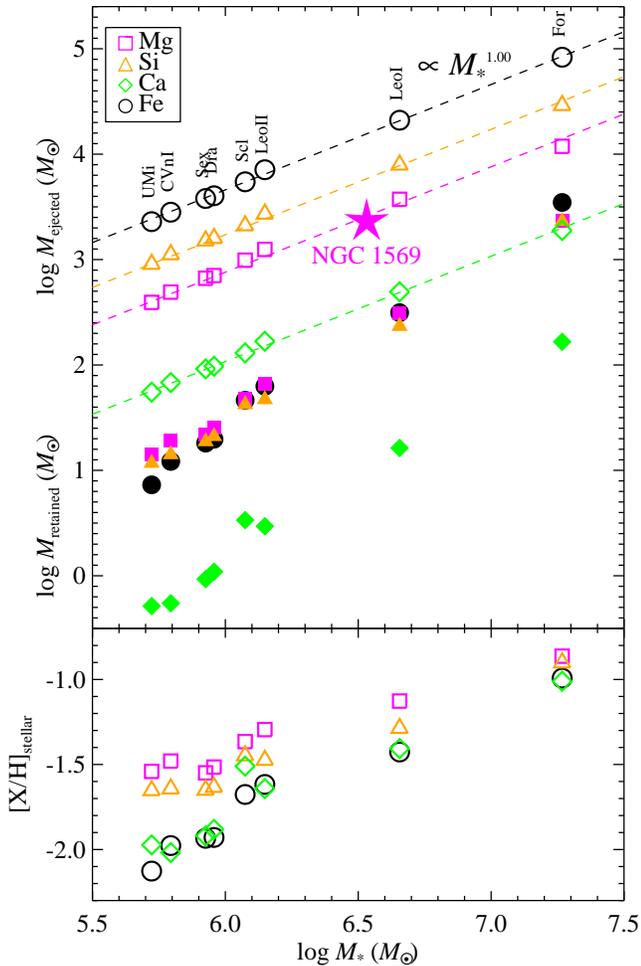}
  \caption{Top: The mass in four different metals inferred to have
    been lost from (hollow points) or retained in (filled points) MW
    dSphs as a function of their present stellar mass.  The hollow
    points show the ``missing metals,'' which should have been
    produced by the SNe in these dSphs but are not present in the
    stars.  The masses of ejected metals include the ejecta of Type~Ia
    SNe that exploded up to 10~Gyr after the cessation of star
    formation.  The dashed lines illustrate slopes of $M_*^{1.00}$,
    which is the expected relation if dSphs lost all of the metals
    they produced.  The large star shows the Mg mass in the current
    outflow from NGC~1569 versus the stellar mass of the most recent
    starburst \protect \citep{mar02}.  Bottom: The stellar
    mass-metallicity relation for the stars in dSphs \protect
    \citep{kir11b}.  Metallicity is given separately as [Fe/H],
          [Mg/H], [Si/H], and [Ca/H].\label{fig:ejecta}}
\end{figure}

\begin{figure}[t!]
  \centering
  \includegraphics[width=\columnwidth]{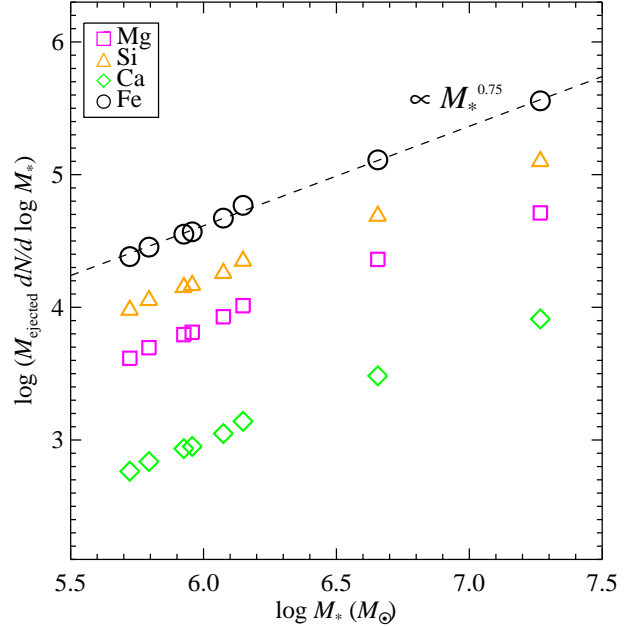}
  \caption{The metals ejected from dSphs convolved with the luminosity
    function for MW dSphs \citep[$dN/d\log M_* \propto
      M_*^{-0.25}$,][]{kop08}.  The dashed line shows $M_*^{0.75}$.
    The axes are logarithmic, and the metal output from small dSphs is
    negligible compared to large dSphs.\label{fig:massfunction}}
\end{figure}

Figure~\ref{fig:massfunction} shows the product of this dSph mass
function and the ejected mass function from Figure~\ref{fig:ejecta}.
Even though the small dSphs are more numerous, they are not
significant contributors to IGM metals.  Even if the relation for the
smaller dSphs is extrapolated to zero mass, then all dSphs the size of
Leo~I or smaller contributed just $7.5 \times 10^4~M_{\sun}$ of Fe to
the IGM over 10~Gyr.  Fornax alone provided more Fe, and even more
massive galaxies likely make Fornax's contribution negligible.  The
same argument applies to the other metals as well.

\subsection{[$\alpha$/Fe]}

Averaged over a long enough time, the ratio of the numbers of Type~II
to Type~Ia SNe asymptotes to the same value in all dSphs as long as
the IMF is the same.  As a result, the [$\alpha$/Fe] ratios of ejecta
from dSphs integrated over 10~Gyr are not functions of stellar mass.
Expressed relative to the solar ratios, the 10~Gyr ratios for our
adopted SN yields and our adopted IMF are $\rm{[Mg/Fe]} = -0.45$,
$\rm{[Si/Fe]} = -0.15$, and $\rm{[Ca/Fe]} = -0.30$.


\section{Comparison to NGC~1569}
\label{sec:ngc1569}

\citet{may06} theorized that gas-rich dIrrs can be converted into
gas-free dSphs by interaction with their host galaxies.  Tidal
stirring can turn disks into spheroids, and ram pressure
stripping---in addition to supernova-driven outflow---can remove all
of the gas.  With this relationship between dIrrs and dSphs in mind,
we compare our inferences for the metals ejected from dSphs to the
observed metal content of the wind presently flowing out the
post-starburst dIrr NGC~1569 \citep[$M_* = 1.8 \times
  10^8~M_{\sun}$,][]{isr88}.

\citet{mar02} estimated that the most recent starburst in that galaxy
ejected $M_{\rm O} \approx 3.4 \times 10^4~M_{\sun}$ of oxygen.  They
based this estimate on an X-ray spectral fit to several $\alpha$
elements, including O and Mg.  In order to compare to our own ejected
metal estimates, we note that $M_{\rm O}$ is related to $M_{\rm Mg}$
by a multiplicative factor (about 15, which is also the solar ratio)
because both elements are formed in similar nucleosynthetic processes
in massive stars.  Therefore, the starburst must have ejected $M_{\rm
  Mg} \approx 2.3 \times 10^3~M_{\sun}$ of magnesium.  The mass of
stars formed in the burst was $M_* \approx 3.4 \times 10^6~M_{\sun}$
over 10--20~Myr.

NGC~1569 is more massive than any of the dSphs we consider, and the
SFR in the burst was much higher than the average rates in the dSphs.
Nonetheless, the outflowing magnesium mass from the recent burst in
NGC~1569 agrees remarkably well with the relation between $M_{\rm Mg}$
and $M_*$ defined by MW dSphs.  Figure~\ref{fig:ejecta} shows that the
NGC~1569 outflow falls exactly in line between Leo~I and Leo~II.  Of
course, we must keep in mind that we are considering only the very
recent starburst and not the entire galaxy, which is 50~times more
massive.  However, only the ejecta of the current starburst is
directly measurable, and the measurement is uncertain by a factor of
at least 2.

\citet{mar02} also measured the [$\alpha$/Fe] ratio to be 0.3--0.6
(2--4 times solar).  This value is larger than what we infer for the
MW dSphs.  However, the winds flowing out of NGC~1569 came from a very
short burst---likely short enough that Type~Ia SN ejecta is not
present in the winds.  Therefore, the [$\alpha$/Fe] ratios merely
reflect the fact that the star formation was intense in the recent
burst in NGC~1569 but not in the dSphs.

\citeauthor{mar02}'s completely independent method for measuring the
metal content of galactic winds has placed a different, though
possibly related, type of dwarf galaxy on the same $M_{\rm
  ejected}$--$M_*$ relation as the one defined by MW dSphs.  This
result has several implications.  First, NGC~1569 retained almost none
of the metals produced in its recent starburst, a result that
\citeauthor{mar02}\ found from their own chemical evolution model.
Second, the theoretical SN yields are accurate enough to place the
observed magnesium mass ejected from NGC~1569 on the same relation as
the the masses ejected from dSphs, which were calculated from the
theoretical yields.  Finally, even a galaxy as massive as NGC~1569
\citep[$\sigma_v = 21$~km~s$^{-1}$,][compared to 10~km~s$^{-1}$ for a
  typical dSph]{sti02} loses nearly all of the metals it produces.


\section{Discussion}

The possibility that the majority of metals that currently reside in
the IGM were originally ejected by low-mass systems \citep{opp08,
  mar10} emphasizes the need for constraints on models for galactic
outflows.  Galaxy evolution models \citep[e.g.,][]{som01,opp06}
generally assume that star-forming galaxies drive outflows whose
metallicity tracks the interstellar medium, and whose outflowing mass
is proportional to $M_*/\sigma^\alpha$.  Here, $\sigma$ is the
velocity dispersion that scales with the halo mass ($\sigma \propto
M_h^{1/3}$), and $\alpha$ is a scaling parameter.  Setting $\alpha=1$
yields a momentum-driven scaling, and $\alpha=2$ yields a
energy-driven scaling.  Some models \citep[e.g.,][]{bow06} invoke a
stronger scaling ($\alpha=3$) in order to match the faint end of the
luminosity function.  Let us assume that the metallicity of outflowing
gas $Z_W$ tracks the stellar metallicity $Z_*$, which in turn scales
with stellar mass as $M_*^{0.31}$ \citep{kir11b}.  Let us further
assume the stellar mass--halo mass scaling that is observed at low
masses \citep[$M_* \propto M_h^{2.17}$,][]{lea11}.  The mass in
ejected metals is the outflow mass multiplied by the wind metallicity.
Therefore, the model assumptions correspond to the assumption that
$M_{\rm ejected} \propto (M_*/\sigma^\alpha)Z_W =
M_*^{1.31}/M_h^{\alpha/3} = M_*^{1.31-\alpha/6.51}$, or $M_*^{1.16}$
and $M_*^{1.00}$ for momentum-driven and energy-driven scalings,
respectively.  Taken at face value, our observation of the ejected
mass relation, $M_{\rm ejected} \propto M_*^{1.00}$, favors an
energy-driven scaling.

We emphasize that these inferences depend strongly on the assumed
stellar mass--halo mass relation as well as the metallicities within
the outflows.  They also neglect many physical processes, such as
tidal stripping, that could preferentially affect low-mass systems.
Furthermore, much of the metals from Type~Ia SNe are lost well after
the cessation of star formation and well after the galaxies are devoid
of interstellar gas.  That makes the distinction between energy- and
momentum-driven winds poorly defined.  These uncertainties emphasize
the need for improved observational constraints on the kinematics of
low-mass systems as well as the mass-loading and metal mass fractions
of outflows from low-mass galaxies.

The biggest limitation of our approach at measuring $M_{\rm ejected}$
is that we have measured the mass only in metals.  We provide no
constraints on the total outflowing mass, including hydrogen.
Therefore, we also provide no constraints on the metal fraction of the
outflows.  One way to constrain the metallicity is to fix the total
mass participating in gas flows and star formation.  For instance, we
could assume that the baryonic mass associated with a dSph obeys the
cosmic baryon-to-dark matter ratio.  Unfortunately, the disconnect
between stellar and dark matter mass \citep{mat98,str08} and the tidal
interactions \citep{pen08} of MW dSphs make it impossible to estimate
their total, original dark matter masses.

Semi-analytic models \citep[e.g.,][]{coo10} may provide a path
forward.  By treating a variety of processes that likely affected dSph
growth, such as tidal stripping, pre-enriched inflows, and returning
outflows, they provide a flexible but complete account of how outflows
impact metallicities.  A model galaxy whose chemical abundances match
observed abundance distributions \citep{kir11a,kir11b} and our
inferred ejected metal masses might inform us about these processes.
We encourage interested modelers to interrogate their existing
semi-analytic models for unique constraints that our observations
place on the way in which low-mass galaxies process their gas into
outflows, stars, and evolving gas reservoirs.

\acknowledgments We thank J.~Cohen and the anonymous referee for
careful, constructive criticism that improved this letter.  Support
for this work was provided by NASA through Hubble Fellowship grants
51256.01 awarded to ENK and 51254.01 awarded to KF by the Space
Telescope Science Institute, which is operated by the Association of
Universities for Research in Astronomy, Inc., for NASA, under contract
NAS 5-26555.  Support to CLM was provided through NSF grant
AST-080816.


{\it Facility:} \facility{Keck:II (DEIMOS)}

\end{document}